\documentclass[useAMS,usenatbib]{mnras}
\usepackage[fleqn]{amsmath}
\pdfoutput=1                    
\usepackage{graphicx}	
\usepackage{multirow}

\newcommand{\der}{{\rm d}}

 \newcommand{\cc}{_{\rm c}}

\newcommand{\pk}{_{\rm pk}} \newcommand{\nest}{^{\rm nest}}
  
\newcommand{\fnest}{^{\rm d\,nest}} \newcommand{\h}{_{\rm h}}

\newcommand{\clM}{M_{\rm s}} 
  
\newcommand{\clS}{S_{\rm s}}

 \newcommand{\dc}{_{\rm dDM}} 
\newcommand{\F}{^{\rm th}} \newcommand{\res}{_{\rm min}} 
 
\newcommand{\p}{_{\rm p}}

\newcommand{\modot}{M$_\odot$\ } \newcommand{\modotc}{M$_\odot$}
\newcommand{\ti}{t_{\rm i}}

\newcommand{\beq}{\begin{equation}} \newcommand{\eeq}{\end{equation}}
 \newcommand{\beqa}{\begin{eqnarray}}
\newcommand{\eeqa}{\end{eqnarray}} \newcommand{\lav}{\langle}
\newcommand{\rav}{\rangle}

\begin{document}

\title[I. Accreted Subhaloes]{An Accurate Comprehensive Approach to Substructure:\\ I. Accreted Subhaloes}

\author[Salvador-Sol\'e, Manrique \& Botella]{Eduard
  Salvador-Sol\'e$^1$\thanks{E-mail: e.salvador@ub.edu}, Alberto Manrique$^1$ and Ignacio Botella$^{1,2}$
  \\$^1$Institut de Ci\`encies del Cosmos, Universitat de
  Barcelona, Mart{\'\i} i Franqu\`es 1, E-08028 Barcelona, Spain
  \\$^2$Dept. of Astronomy, Graduate School of Science, Kyoto University, Kitashirakawa, Oiwakecho, Sakyo-ku, Kyoto, 606-8502, Japan}


\maketitle
\begin{abstract}
This is the first of a series of three Papers devoted to the study of halo substructure in hierarchical cosmologies by means of the CUSP formalism. In the present Paper we derive the properties of subhaloes and diffuse dark matter (dDM) accreted onto haloes and their progenitors. Specifically, we relate the dDM present at any time in the inter-halo medium of the real Universe or a cosmological simulation with the corresponding free-streaming mass or the halo resolution mass, respectively, and establish the link between subhaloes and their seeds in the initial density field. By monitoring the collapse and virialisation of haloes, we derive from first principles and with no single free parameter the abundance and radial distribution of dDM and subhaloes accreted onto them. Our predictions are in excellent agreement with the results of simulations, but for the predicted fraction of accreted dDM, which is larger than reported in previous works as they only count the dDM accreted onto the final halo, not onto its progenitors. The derivation pursued here clarifies the origin of some key features of substructure. Overall, our results demonstrate that CUSP is a powerful tool for understanding halo substructure and extending the results of simulations to haloes with arbitrary masses, redshifts and formation times in any hierarchical cosmology endowed with random Gaussian density perturbations. 
\end{abstract}



\section{INTRODUCTION}\label{intro}

Over the last two decades halo substructure has become a subject of paramount importance as it plays a crucial role in many astrophysical issues. This is the reason that great effort has gone into accurately determining the typical properties of dark matter halo substructure.

This has been mostly achieved by means of high-resolution $N$-body simulations. After some preliminary works \citep{Gea98,Tea98,Sea,Helmi02,Stea03,Gae04,Dea04b,Kea04,DeL04,Rea5}, convergence was reached and very detailed studies were carried out on Milky Way (MW)-mass haloes in the $\Lambda$CDM cosmology (\citealt{Dea07} and \citealt{Sea08a}, hereafter SWV; \citealt{Dea08,Gi08}). Since then $N$-body simulations have been applied to investigate substructure in haloes of different masses and redshifts in CDM (\citealt{Lea09,Aea09,Eea09,Gi10,Kea11,Gea11,Oea12,Gea12,Kea13,Cea14,Gfea16,Hea18}) as well as WDM \citep{Lea14} cosmologies and, in particular, focusing on the effects on gravitational lensing (e.g. \citealt{MM06,Xea10,Cea11}). Recently, simulations have included baryons in order to study dwarf satellites in MW analogs (\citealt{Hell16,Bea16,Bea20,Rich20,Fea20,Fea20b}).

But high-resolution simulations are very CPU-expensive and only cover a limited subhalo mass range in massive haloes at low redshifts and in a few cosmologies; they are not free of numerical effects \citep{Cea14,vdB18,vdBO18}; and, even though they are well-suited to characterise the typical properties of substructure, they are not to understand their origin.

This has led to explore alternative approaches. Apart from the incipient use of machine learning techniques applied to gravitational lensing observations (e.g. \citealt{Aea20,Vea20}), many studies have used numerical experiments or analytic models. The modelling of substructure is very complex, however, and requires making important simplifications (e.g. \citealt{Fea02,S03,L04,OL04}). Very complete and detailed models based on the extended Press-Schechter formalism \citep{PS,B91,BCEK,LC93} have been built, though at the expense of introducing numerous free parameters tuned through the comparison with simulations \citep{TB01,ZB03,TB04,Pe05,vdB05,Zea05,KB07,Bea13,PB14,Jv16,vdB16}. It is thus hard to tell to what extent the good results obtained by these models are not due to that large freedom. In addition, they do not clarify the link between the final properties of substructure and each individual process at work. This is the reason that some authors have preferred to concentrate in specific aspects of the problem such as the role of stripping or disruption (e.g. \citealt{vdBbis16,Jv17,vdB18,vdBO18}).

An important step along the modelling of substructure was taken by Han et al. (2016; hereafter HCFJ). These authors showed that the observed properties of substructure follow from three only conditions: 1) the scaled number density profiles of subhaloes with original mass $\clM$ accreted by the halo or its progenitors at all previous times overlap in one curve proportional to the scaled density profile of the host halo; 2) the cumulative mass function (MF) of accreted subhaloes is a power-law with logarithmic slope, $\der {\cal N}(> \clM)/\der \ln \clM$, close to $-1$; and 3) the truncated-to-original mass ratio of the final stripped subhaloes only depends on their radial distance to the centre of the host halo. Strictly speaking, the two former conditions refer to `unevolved' subhaloes (i.e. at their final radius but with their original mass at the infall time) rather than to `accreted' ones (i.e. with their mass and radius at accretion). But that difference should only be relevant for massive subhaloes which suffer dynamical friction (see below).\footnote{We have implicitly assumed here that the infall time of subhaloes coincides with what in the present Paper is defined as their accretion time, which may not be the case, however (see Sec.~\ref{accreted}).}

Unfortunately, the way HCFJ conditions are set is unknown, so the ultimate origin of the observed properties of substructure remains to be elucidated. The aim of the present Paper and two following ones (\citealt{II} and \citealt{III}, hereafter Papers II and III) is to shed light on that issue through a new comprehensive approach making use of the so-called {\it ConflUent System of Peak trajectories} (CUSP) formalism \citep{Mea95,Mea96,Mea98}. 

CUSP makes the link between haloes and their seeds, peaks (or maxima), in the primordial Gaussian random field of density perturbations in any given hierarchical cosmology \citep{Jea14a}. Thus, following the collapse and virialisation of those seeds, with well-known properties according to Gaussian statistics, one can accurately derive {\it from first principles and with no single free parameter} the properties of virialised haloes.\footnote{The virial relation we refer to throughout the Paper includes the external pressure term. Thus, by virialised haloes we simply mean haloes in equilibrium.} The halo properties so derived, namely the MF \citep{Jea14b}, halo inner structure (\citealt{Sea12a}, hereafter SVMS), kinematics and triaxial shape (\citealt{Sea12b}, hereafter SSMG) fully agree with the results of $N$-body simulations both in CDM (all preceding references) and WDM \citep{Vea12} cosmologies. More importantly, their derivation clarifies the origin of their characteristic features (see \citealt{SM19} for an overview). 

In the series of three Paper initiated here we extend the application of CUSP to the study of halo substructure. Specifically, here we relate the amount of diffuse dark matter (dDM) present at any given time in the inter-halo medium with the free-streaming mass or halo resolution mass in the real Universe or the cosmological simulation, respectively, and establish the link between subhaloes and their seeds in the initial density field. This link allows us to derive, again from first principles and with no single free parameter, the radial distribution and MF of accreted subhaloes. The derivation clarifies the origin of the two first HCFJ conditions. In Papers II and III we address the more complex problem of the fate of subhaloes orbiting inside haloes and its dependence on the halo assembly history and clarify the origin of HCFJ condition 3.

The layout of the Paper is as follows. In Section \ref{peaks}, we remind the CUSP formalism. In Section \ref{nesting}, we extend it so as to deal with substructure. The radial distribution and MF of accreted subhaloes are derived in Section \ref{accreted}. The limits of the approach are discussed in Section \ref{friction} and a summary of our results is given in Section \ref{dis}.

Throughout the Paper our predictions are calculated for current Milky Way (MW)-like haloes with virial mass $M\h$, i.e. the mass out to the radius encompassing an inner mean density equal to the virial overdensity \citep{BN98,H00} times the mean cosmic density, of $M\h=2.2\times 10^{12}$ \modotc, which according to \citet{SM19} correspond to the maximum extend of the virialised part of haloes. These predictions are compared to the results for the same kinds of haloes studied by HCFJ, who use the masses $M_{200}$, i.e. the mass out to the radius encompassing an inner mean density of 200 times the critical cosmic density, of $1.84\times 10^{12}$ \modotc. The cosmology adopted is the {\it WMAP7} cosmology \citep{Km11} as in those latter works. The CDM spectrum we use is according to the prescriptiobn given by \citet{BBKS} (hereafter BBKS) with the \citet{S95} shape parameter.

\section{The CUSP Formalism}\label{peaks}

CUSP is a formalism that monitors the halo clustering process through the filtering of the random Gaussian field of dark matter density perturbations at an arbitrary initial time $\ti$. It is thus similar to the excursion set (ES) formalism \citep{BCEK}. The main differences between the two are: 1) while the ES formalism looks at the changes in the density contrast $\delta$ {\it at fixed points} produced when the scale $S$ of the smoothing filter is varied, CUSP looks at the changes in $\delta$ at {\it moving peaks}; 2) while the ES formalism uses a $k$-sharp smoothing window, causing the $\delta(S)$ trajectories to follow random walks, CUSP uses a Gaussian window, which guarantees that $S$ increases with decreasing $\delta(S)$ just as halo masses increase with increasing time; 3) while the ES formalism uses the top-hat spherical (or ellipsoidal) collapse model to determine the collapse time of patches of different masses found by means of the $k$-sharp window, \footnote{The ellipsoidal collapse is dealt in an approximate way through a varying critical density for collapse governed by three parameters \citep{Sea01}.} CUSP uses for this purpose the same Gaussian window used to find the collapsing patches; and 4) while ES is not fine enough to distinguish between different halo mass definitions, CUSP is.

But the most notable characteristic of CUSP is that, not only does it provide, like the ES, the MF of haloes collapsed any time, but it also allows one to accurately derive all their macroscopic properties.

We remind next the main results of CUSP used in the present work (see \citealt{SM19} for a more complete and detailed overview).

\subsection{Haloes and Peaks}

\begin{itemize}

\item{\it Halo-peak correspondence:} There is a one-to-one correspondence between virialised haloes with mass $M$ at the cosmic time $t$ and {\it non-nested} peaks with density contrast $\delta$ at the scale $S$ in the Gaussian-filtered Gaussian random density field at an arbitrary initial time $\ti$ given by \citep{Jea14a}
\beq 
\delta(t)=\delta\cc(t)\frac{D(\ti)}{D(t)}\,
\label{deltat}
\eeq
\beq 
\sigma_0^2(M,t)= \frac{1}{2\pi^2}\int_0^\infty \der k\,k^2
P(k,t)\,\exp\left(-k^2S^2\right),
\label{rm}
\eeq
were $\delta\cc(t)$ is the Gaussian-filtered critical linear density contrast for {\it ellipsoidal collapse at $t$}, $D(t)$ is the cosmic growth factor, $P(k,t)$ is the (linear) power spectrum of density perturbations at $t$ and $\sigma_{\rm j}(M,t)$ are the corresponding jth order Gaussian spectral moments. 

$\delta\cc(t)$ and $\sigma_0(M,t)$ can be approximately written in terms of their more usual top-hat counterparts for spherical collapse, $\delta\cc\F(t)$ and $\sigma_0\F(M,t)$, as
\beqa
\delta\cc(t)\approx\delta\cc\F(t)\frac{a^d(t)}{D(t)}~~~~~~~~~~~~~~~~~~~~~~~~~~~~~~~~~~~~~~~~~~~~~~~\\
\sigma_0(M,t)
\label{delta0}
\!\approx\!\sigma_0\F(M,t)\!+\!\left\{\!s_0+s_1a(t)+\log\!\left[\!\frac{Aa^{s}(t)}{A+a(t)}\!\right]\!\right\}\!\delta(t),\!\!\!\!
\label{sigma}
\eeqa
where $a(t)$ is the cosmic scale factor and coefficients $d$, $s_0$, $s_1$, $s$ and $A$ depend on cosmology and halo mass-definition. For those considered in the present work in order to compare with HCFJ results, they are respectively equal to 1.06, 0.0422, 0.0375, 0.0318 and 25.7. Note that for such values of the parameters $\sigma_0(M,t)$ is close to $\sigma_0\F(M,t)$, implying that $M$ is close to $4\pi S^3\rho\cc(\ti)$, where $\rho\cc(\ti)$ is the mean cosmic density at $\ti$. 

\item{\it Peak number density and halo MF:} 
The comoving mean number density of peaks with $\delta$ per infinitesimal scale around $S$ at $\ti$ \citep{Mea95} is
\beq
N\pk(S,\delta)=\frac{\lav x\rav(S,\delta)}{(2\pi)^2S_\star^3}\,\,
\exp\left(-\frac{\nu^2}{2}\right)\,
{\frac{\sigma_2(S)}{\sigma_0(S)}}
\,S\,,
\label{npeak}
\eeq
where $\nu$ is the peak height, $\delta/\sigma_0(S)$, $S_\star$ is defined as $\sqrt{3}\sigma_1(S)/\sigma_2(S)$, and $\lav x\rav(S,\delta)$ is the mean curvature of peaks with $\delta$ at $S$ (BBKS). For moderately high peaks, as it corresponds to haloes of galactic scales, $\lav x\rav(S,\delta)$ is well-approximated by $\gamma \nu$,
where $\gamma$ is $\sigma_1^2/(\sigma_0\sigma_2)$, so it very nearly factorises as a function of $\delta$ times another function of $S$. Thus, the number density $N\pk(S,\delta)$ also does. 

This mean number density of peaks must be corrected, however, for nesting because some peaks with $\delta$ at $S$ are nested within other peaks with $\delta$ at $S'>S$ and will not give rise to virialised haloes, but to subhaloes within other virialised haloes (see below). This correction is achieved by solving the Volterra integral equation of second kind \citep{Mea95},
\beqa
N(S,\delta)=N\pk(S,\delta)\nonumber~~~~~~~~~~~~~~~~~~~~~~~~~~~~~~~~~~~~~~~~~~~~~\\
-\int_S^\infty dS' N\pk\nest(S,\delta|S',\delta)\, N(S',\delta) \frac{M(S')}{\rho\cc(\ti)}\,,~~~
\label{nnp}
\eeqa
where $N\pk\nest(S,\delta|S',\delta)$ is the mean conditional number density of peaks with $\delta$ per infinitesimal scale around $S$ subject to being nested in non-nested peaks with $\delta$ at $S'$, equal to the integral over $r$,
\beq 
N\pk\nest(S,\delta|S',\delta)= C\int_0^1\!
\der r\, 3 r^2 N\pk(S,\delta|S',\delta,r)\,,
\label{int}
\eeq
of the same quantity subject to the additional condition of being at a distance $r$ (in units of the scale $S'$) from the host peak \citep{Mea98}, 
\beqa 
N\pk(S,\delta|S',\delta',r)~~~~~~~~~~~~~~~~~~~~~~~~~~~~~~~~~~~~~~~~~~~~~~~~\nonumber\\
=\frac{\lav x\rav(S,\delta,S',r)}{(2\pi)^2\,S_{\star}^3\,e(r)}\exp\left\{\!-\frac{\left[\nu -
\epsilon(r)\,\nu'(r)\right]^2}{2e(r)^2}\!\right\}{\frac{\sigma_2(S)}{\sigma_0(S)}}\,S\,,
\label{mf16}
\eeqa 
where $\lav x\rav(S,\delta,S',r)$ is the mean curvature of peaks with $\delta$ and $S$ at a distance $r$ from another peak with $\delta$ and $S'$, and $\nu'(r)=\overline{\delta(r)}/\sigma_0(S') g(r,S')$, $e(r)=\sqrt{1 - \epsilon^2(r)}$, $\epsilon(r)=\sigma_0^2(S_{\rm m})/[\sigma_0(S)\sigma_0(S')] g(r,S')$, $S^2_{\rm m}=SS'$, and $g(r,S')=\sqrt{1-[\Delta\delta(r)]^2/\sigma_0^2(S')}$, where $\overline{\delta(r)}$ and $\Delta\delta(r)$ are the mean and rms density contrast, respectively, at $r$ from the host peak (BBKS). Factor $C$ in equation (\ref{int}), given by
\beq
C\equiv\frac{4\pi S^3 N\pk(S',\delta)}{3N\pk(S,\delta)}
\int_0^{L}\der r\,3 r^2\,N\pk(S,\delta|S',\delta,r)\,,
\label{C}
\eeq
where $L$ is the mean non-nested peak separation calculated from their mean density (eq.~[\ref{nnp}]), is to correct the host peaks themselves for nesting. 

Similarly to $\lav x\rav(S,\delta)$, the mean curvature $\lav x\rav(S,\delta,S',r)$ in equation (\ref{mf16}) very nearly factorises as a function of $S$ times a function of the remaining arguments, $\delta$, $S'$ and $r$. Hence, in the relevant subhalo mass regime where the exponential on the right of equation (\ref{mf16}) is approximately unity, $N\pk(S,\delta|S',\delta,r)$ also factorises in the same way, which implies in turn that $C$ (eq.~[\ref{C}]) and $N\pk\nest(S,\delta|S',\delta)$ (eq.~[\ref{int}]) do, too.

Given the one-to-one correspondence between virialised haloes and non-nested peaks, by changing $\delta$ to $t$ and $S$ to $M$ (eqs.~[\ref{deltat}] and [\ref{rm}]), the comoving halo MF at $t$ gives the comoving density of non-nested peaks given by the solution of equation (\ref{nnp}).

\item{\it Unconvolved inner peak properties:} The continuous mean peak trajectory $\delta(S)$ for haloes with $M\h$ at $t\h$ having been accreting at the mean instantaneous rate $\der M/\der t$ is the solution for $\delta(t\h)$ at $S(M\h,t\h)$ of the differential equation (SVMS)
\beq 
\frac{\der \delta}{\der S}=-\lav x\rav(S,\delta)\,\sigma_2(S) S\,.
\label{dmd}
\eeq
where $\delta(S)$ is the mean density contrast of the corresponding peaks at scale $S$. This continuous trajectory is the convolution, with a Gaussian window of running scale $S$, of the mean unconvolved spherically averaged density contrast profile $\delta\p({\bf r})$ around the peak centre. Thus, it can be inverted (SVMS) to obtain the mean spherically averaged density profile $\rho\p({\bf r})$ around peaks with given $\delta$ at $S$. Likewise, one can derive the mean unconvolved spherically averaged ellipticity and prolateness profiles around peaks (SSMG).

\item{\it Inner halo properties:} 
The unconvolved inner profiles of peaks associated with purely accreting haloes can be used to derived the inner properties of such haloes.

Indeed, after expanding radially in the initial linear regime, ellipsoidal shells (peaks are triaxial) reach turn-around, collapse non-radially and bounce. The shell-crossing produced is what leads to the virialisation of the system. The way this is achieved is easily seen in the spherical symmetry approximation. When a shell moving outwards crosses another shell moving inwards, some amount of gravitational energy is transferred between them, which depends on the radius where the crossing takes place: near pericentre they exchange more energy than near apocentre (see SVMS for details). This difference leads to a continuous energy outflow from shells collapsing earlier to shells collapsing later, causing shell orbits to orderly shrink. The increasingly disordered shell-crossing progressively erases the initial correlation between the orbital phases of different shells, which reduces the energy outflow outwards. When the correlation fully disappears shells stop contracting. The earlier a shell is accreted (and the smaller is its turnaround radius), the earlier it stabilises. And as, despite shell-crossing, there is no apocentre-crossing (SVMS) during that `gentle relaxation', {\it accreting haloes grow inside-out}. 
  
Strictly speaking, the orbits of particles in the same ellipsoidal shell at turnaround slightly differ from each other and the apocentres of objects in neighbouring shells slightly mix in the final virialised halo. But this does not alter the main conclusion that haloes essentially grow inside-out.   

As objects orbiting within haloes spend most of the time near their apocentre, most of the objects caught at any radius $r$ are at their stabilised apocentre, implying that they were accreted at the time $t(r)$ when the mass of the halo inside $r$ was $M(r)$. We are then led to the following approximate relation between the radius $r$ and the mass $M$ within it 
\beq 
r=\frac{10}{3}\frac{GM^2}{|{\cal E}\p(M)|}\,,
\label{vir2}
\eeq
where ${\cal E}\p(M)$ is the (non-conserved) total energy in the sphere of mass $M$ around the (triaxial) peak and $G$ is the gravitational constant. ${\cal E}\p(M)$ can be calculated from the known mean spherically averaged density profile of their seeds and the perturbed Hubble-flow velocity field around them (SVMG). Equation (\ref{vir2}) leads to the mean mass profile $M(r)$ and, by differentiation, to the mean spherically averaged density profile $\rho(r)$ for purely accreting haloes with $M\h$ at $t\h$.

Similarly, one can derive the mean spherically averaged ellipticity and prolateness profiles together with the mean spherically averaged velocity dispersion and anisotropy profiles for purely accreting haloes from the unconvolved inner ellipticity and prolateness profiles of their seeds (SSMG).  
\end{itemize}

It is important to remark that, even though in the previous derivations we have concentrated in haloes evolving by pure accretion (monolithic collapse), haloes also undergo major mergers (lumpy collapse). The CUSP formalism makes the distinction between accretion and major mergers in a rigorous way. The reader is referred to \citet{SM19} for a detailed explanation of such a distinction and its consequences on peaks. We will just mention here that when a halo accretes other (less massive) haloes, the latter haloes are not destroyed and become subhaloes of the former, while when two (similarly massive) haloes undergo a major merger they are destroyed (they do not survive as subhaloes) but their own subhaloes become subhaloes of the new halo formed in the merger. This process of substructure growth is truthfully followed by the evolution of peaks in the $\delta$-$S$ plane, which develop in parallel a complex nesting structure. Taking profit of that parallelism, we can study the growth of halo substructure from the statistics of peak nesting. 

Note that, even though the first generation haloes do not harbour subhaloes because they from by monolithic collapse of the diffuse DM (dDM) prior to halo formation, the second generation haloes begin to accrete first generation ones which become subhaloes, which marks the beginning of the process of substructure growth. (If there were only major mergers, the second generation haloes would keep on being substructure-free and substructure would never form.) Thus, the distinction between accretion and major mergers is crucial for substructure. Nevertheless, \citet{SM19} have shown that all macroscopic halo properties, including their content of subhaloes and dDM, do not depend on their individual assembly history (but see Papers II and III for the consequences of stripping). Thus, one has the right to assume in the present study that haloes grow by pure accretion with no loss of generality. We will only refer to major mergers at the end of next Section when comparing our predictions on accreted dDM with the results of some simulations where the only dDM counted is that accreted by haloes ``since their last major merger".

\subsection{Diffuse Dark Matter}\label{dDM}

When correcting for nested configurations (eq.~[\ref{nnp}]), we have implicitly assumed that, at any time $t$, all the dark matter present in the Universe is locked within virialised haloes of different masses. Indeed, in all hierarchical cosmologies, the first haloes to form are those of the smallest mass, which then progressively grow through accretion and major mergers. This means that the power spectrum of density perturbations is such that the variance and, hence, the number density of peaks always increases with decreasing scale. In other words, unless there is a lower bound in the scale (a cutoff in the power spectrum) the number density of peaks will diverge at the small scale end. Thus, in CDM cosmologies it is natural to assume that haloes cover the whole cosmic dark matter. 

However, in the real Universe as well as in cosmological simulations, haloes can only form with masses above a minimum value, $M\res$, related with the free-streaming mass associated with the WIMP or the halo resolution mass used in the simulation. Consequently, at any $t$ greater than the starting time for halo clustering, $t\res$, all the dark matter that would lie in haloes of masses below $M\res$ will remain in the diffuse form in the inter-halo medium. Of course, such a dDM tends to vanish as it is progressively accreted onto massive haloes, just as the corresponding idealised low-mass haloes would be.

\begin{figure}
\centerline{\includegraphics[scale=.60,bb= 18 40 500 500]{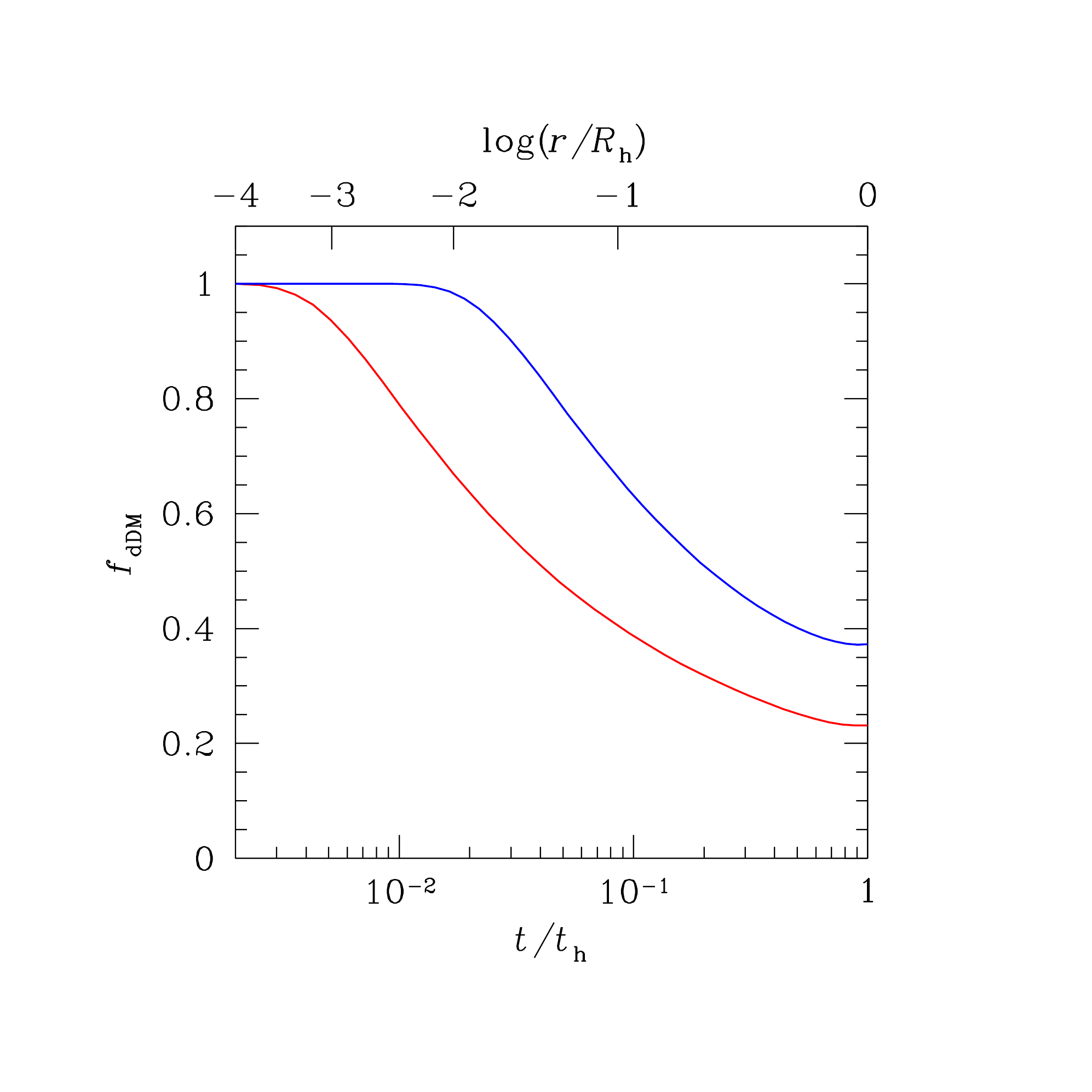}}
\caption{Diffuse dark matter mass fraction in the inter-halo medium as a function of cosmic time in a real 100 GeV WIMP universe (red colour) and an Aquarius-like simulation (blue colour) as a function of cosmic time (lower tic-labels). The same curves also give the dDM mass fraction at the radius $r$ of a purely accreting halo, i.e. growing inside-out, with current MW mass.}
(A colour version of this Figure is available in the online journal.)
\label{f1}
\end{figure}

According to this picture, the evolving dDM mass fraction $f\dc(t)$ in the inter-halo medium at any time $t>t\res$ satisfies the relation
\beq
f\dc(t)=1-\frac{1}{\rho\cc(t)}\int_{M\res}^\infty\der M\, M\,N(M,t),
\label{feq}
\eeq
where $N(M,t)$ is the above derived halo MF at $t$. In Figure \ref{f1} we plot the dDM mass fraction in the inter-halo medium as a function of cosmic time. The two solutions depicted in correspond to: 1) a real 100 GeV WIMP universe with a minimum halo mass $M\res$ of $10^{-6}$ \modot at the time of decoupling, as studied by \citet{AW10}; and 2) an Aquarius Level 2 resolution-like simulation (SWV), which start at $z=127$ ($t=0.0124$ Gyr) with a DM particle mass of $m_{\rm p}=1.37\times 10^4$ \modot and a halo resolution mass of 32 particles, i.e. $M\res=4.4 \times 10^5$ \modotc. (HCFJ concentrate, instead, on the simulation of the A halo reaching the highest (Level 1) resolution, with $M\res\simeq 10^5$ \modotc.)

\begin{figure}
\centerline{\includegraphics[scale=.60,bb= -40 45 580 502]{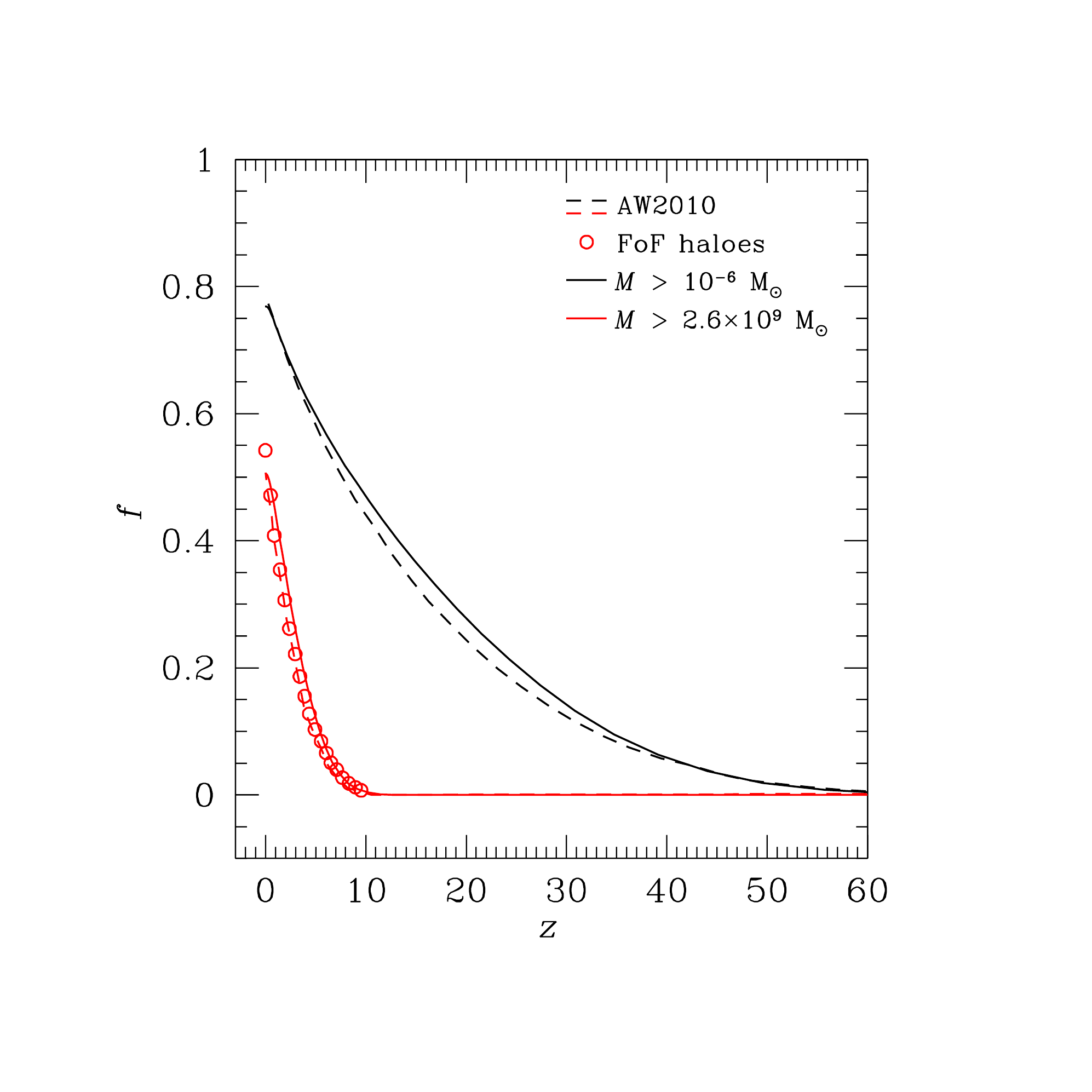}}
\caption{Total cosmic mass fractions in haloes above the two quoted masses as a function of redshift predicted by CUSP in a real 100 GeV WIMP universe (solid lines) compared to the estimates carried out by \citet{AW10} using the ES formalism with ellipsoidal moving barrier (dashed lines) and $N$-body simulations (open circles). Note that both methods predict a current mass fraction of dDM of about 0.23. The very slight deviation of our predictions from the result of simulated haloes can be entirely due to the slightly different cosmology and halo mass definition used in both works.}
(A colour version of this Figure is available in the online journal.)
\label{f2}
\end{figure}

As the amount of dDM in the intra-halo medium diminishes with increasing time, the dDM mass fraction accreted onto haloes diminishes accordingly. Specifically, given that accreting haloes grow inside-out, the fraction $f\dc(r)$ of dDM at the radius $r$ of a halo should be approximately equal to that in the inter-halo medium by the time $t(r)$ when the halo has assembled the mass $M(r)$. In other words, $f\dc(r)=f\dc[t(r)]$ should give the dDM mass fraction at radius $r$ within haloes. That mean spherically averaged dDM mass fraction profile is also plotted in Figure \ref{f1} for MW-mass haloes. But, as $f\dc(t)$ is independent of halo mass and $\rho(r/R\h)$ and $t(r/R\h)$ are roughly universal \citep{SM19}, haloes of all masses should have similar $f\dc(r/R\h)$ profiles. We remark, however, that this profile is only approximate because haloes of different masses do not accrete dDM in exactly the same proportion. Actually, the dDM mass fraction accreted onto haloes at any given time $t$ and, hence, found at $r(t)$ is more exactly given by the mass fraction in virtual accreted subhaloes with masses below $M\res$ (see Sec.~\ref{accreted}). As the more massive a halo, the higher the upper mass of subhaloes it can accrete, the mass fraction in virtual subhaloes with masses below $M\res$ is smaller. In other words, we foresee a slight trend for more massive haloes to have slightly lower dDM mass fractions. Nonetheless, the profile plot in Figure \ref{f2} is a good approximation.


Using the relation (\ref{feq}) we find that, in the WIMP universe, 23\% of the dark matter is currently in the diffuse form and 32\% of the mass of MW-like haloes corresponds to accreted dDM and, in the Aquarius-like simulation, 31\% the dark matter in the Universe is currently in the diffuse form and 51\% of the mass of MW-like haloes corresponds to accreted dDM. How do these values compare to previous estimates? 

\citet{AW10} monitored the evolution of dDM in the 100 GeV WIMP universe through Monte Carlo random walks according to the ES formalism with ellipsoidal moving barrier.\footnote{To avoid the crossing of the barrier at small masses \citet{AW10} took the three parameters entering the ES formalism with slightly different values from those giving the best fit to the halo MF at $z=0$ \citep{ST02}. This illustrates the difficulty to deal in ES framework with the ellipsoidal collapse for haloes spanning more than $20$ orders of magnitude in mass as required in this case.} The study was completed with full $N$-body simulations for haloes having reached high enough masses. The total dDM mass fraction for the Universe at $z=0$ found by these authors is consistent with ours (see Fig.~\ref{f2}). However, they found only $\sim 10$\% of the mass of current MW-like haloes is in dDM, which is notably less than we find.

It is true that the results reported by \citet{AW10} refer to ordinary haloes, i.e. haloes growing by accretion and major mergers, while we have assumed purely accreting haloes. But, as mentioned, our results hold for ordinary haloes as well \citep{SM19}, so this difference does not explain the discrepancy. What does explain it is, precisely, that CUSP accounts for {\it all} the dDM present in haloes regardless of their assembly history, i.e. regardless of whether it has been directly accreted by the final object or by any of its progenitors and brought through major mergers into the main parent halo. Indeed, when two haloes merge, both their subhaloes and their dDM are transferred to the new halo resulting from the merger. Instead, Angulo and White only count the dDM that has been `directly accreted' by the final object. We thus conclude that, in MW-like haloes, typically $\sim 10$\% of their mass is in the form of dDM directly accreted by them, as stated by \citet{AW10}, and another $\sim 20$\% has been accreted by their progenitors and brought into the final object through mergers.

The results found by \citet{Wa11} in the (Level 2 resolution) Aquarius simulations (SWV) support this conclusion. These authors monitored the growth of 6 MW-mass haloes and found values of the dDM mass fraction directly accreted onto them which vary from slightly more than $10$\% to slightly less than $50$\% depending on the exact procedure used to find the accreted dDM and the assembly history of each individual halo. In particular, the A and C haloes, which have suffered their last major merger a long time ago ($z\la 6$), have the largest fractions of accreted dDM near our predicted value because most of their final mass has been assembled through accretion. In contrast, the other four haloes, which have suffered recent major mergers and, hence, have assembled most of their final mass in that way, have the lowest fractions of accreted dDM (less than 40\% and even less than 35\% in the F halo).

\section{SUBHALOES AND NESTED PEAKS}\label{nesting}

When a halo is accreted onto another halo and becomes a subhalo, its corresponding peak becomes nested on the peak associated to the host halo. Despite that nesting, the peak keeps on tracing its continuous trajectory in the $\delta$--$S$ plane \citep{Mea95}. Peak trajectories are only interrupted when their corresponding haloes suffer a major merger. Then, the peaks corresponding to the halo progenitors, with identical density contrast, disappear and a new peak appears with the same density contrast but on a substantially larger scale that initiates a continuous trajectory tracing the new accreting halo formed in the merger \citep{Mea95}. As the peaks nested in the progenitor peaks do not merge themselves, they survive as peaks nested into the new peak, just as subhaloes in the progenitor haloes become subhaloes of the new halo. 

As a consequence, halo substructure at all levels translates into a network of peaks with identical $\delta$ on different scales $S$ nested inside each other at all levels \citep{AJ90}. Moreover, such a peak network varies with running $\delta$ in a way that trustfully traces the evolution of substructure in haloes growing through accretion and major mergers. Therefore, by monitoring the evolution with varying scale of the peak network with known statistics, one can monitor the dynamical evolution of substructure.

This way CUSP can be used to study substructure at any desired level. However, the higher the level, the more complicated the calculations. Fortunately, the general behaviour of substructure is already encoded in the properties of first-level subhaloes, so we will concentrate from now on in {\it first-level} nested peaks.  

When a halo is accreted onto a halo and becomes a subhalo, it begins to be stripped and loses mass, while the associated nested peak keeps on tracing a continuous trajectory as if the subhalo kept on accreting. Consequently, the previous halo-peak correspondence breaks down. However, at the precise moment of the accretion, it still holds, so {\it accreted} subhaloes are traced by their associated nested peaks through equations (\ref{deltat}) and (\ref{rm}) at their respective accretion time. And to derive the properties of substructure regarding accreted subhaloes we must calculate the properties of peaks nested into the larger scale peak tracing the host halo. From now on, peak numbers are denoted with calligraphic $N$ to distinguish them from peak number densities calculated so far.

The mean conditional number of peaks with $\delta$ per infinitesimal scale around $S$ subject to being directly nested in one non-nested peak with $\delta$ at $S'$ is  
\beq 
{\cal N}(S,\delta|S',\delta)=\frac{M(S')}{\rho\cc(\ti)}
N\nest(S,\delta|S',\delta)\,, 
\label{peaknum} 
\eeq 
where $N\nest(S,\delta|S',\delta)$ is the mean conditional number density of peaks with the same characteristics subject to being nested in non-nested peaks with identical $\delta$ at $S'$ (eq.~[\ref{int}]) {\it corrected for nesting at any intermediate scale $S''$ between $S$ and $S'$}, i.e. the solution of the Volterra equation of second kind
\beqa 
N\nest(S,\delta|S',\delta)=
N\pk\nest(S,\delta|S',\delta)~~~~~~~~~~~~~~~~~~~~~~~~~~~\nonumber\\
\!-\!\int_{S}^{S'}\!\!\!\der
S''\,N\pk\fnest(S,\delta|S'',\delta)\,N\pk\nest\!(S'',\delta|S',\delta)\,\frac{M(S'')}
{\rho\cc(\ti)}.\!
\label{peaks2} 
\eeqa
To avoid overcorrection for intermediate nesting (peaks can be nested in more than one intermediate scale peak), in the integral on the right of equation (\ref{peaks2}), we have used the comoving mean conditional number density of peaks with $\delta$ per infinitesimal scale around $S'$ subject to being {\it directly} nested within peaks at $S''$ (i.e. without being nested in any smaller scale peak), $N\pk\fnest(S',\delta|S'',\delta)$, given by the solution of the new Volterra equation
\beqa N\pk\fnest(S,\delta|S',\delta)\equiv
N\pk\nest(S,\delta|S',\delta)
~~~~~~~~~~~~~~~~~~~~~~\nonumber\\ -\!\int_{S}^{S'}\!\!\der
S'' N\pk\fnest\!(S'',\delta|S',\delta)N\pk\nest(S,\delta|S'',\delta)
\frac{M(S'')}
{\rho\cc(\ti)}.\!
\label{cor}
\eeqa

As $N\pk\nest(S,\delta|S',\delta)$ very nearly factorises as a function of $S$ times a function of the remaining arguments, so does $N\pk\fnest(S,\delta|S',\delta)$, as readily seen by dividing equation (\ref{cor}) by the function of $S$ in the former factorisation. Moreover, for identical reason, $N\nest(S,\delta|S',\delta)$ in equation (\ref{peaks2}) and then ${\cal N}(S,\delta|S',\delta)$ in equation (\ref{peaknum}) do too. Thus the mean conditional number of peaks with $\delta$ per infinitesimal scale around $S$ subject to being directly nested in a non-nested peak with $\delta$ at $S'$ is very nearly separable as a function of $S$ and a function of the remaining arguments. As shown below, this separability is crucial for the two first HCFJ conditions.

Given the one-to-one correspondence between subhaloes and nested peaks, we have that the mean number of accreted subhaloes per infinitesimal mass around $\clM$ in haloes with mass $M(r)$ at the time $t(r)$, ${\cal N}[t(r),\clM|t(r),M(r)]$, from now on simply ${\cal N}(<\! r, \clM)$, is given, through the appropriate change of variables, by the mean number of peaks per infinitesimal scale $S(\clM)$ with $\delta[t(r)]$ directly nested into peaks with the same density contrast at the larger scale $S'[M(r)]$. 

We remark that we are not making the assumption that haloes are spherically symmetric. All previous quantities refer to the mass and subhalo number inside the distance $r$ from the centre of mass of the halo or, equivalently, at a radius $r$ of the {\it spherically averaged system} around that centre.    

\section{RADIAL DISTRIBUTION AND MASS FUNCTION OF ACCRETED SUBHALOES}\label{accreted}

As mentioned, even though the mean spherically averaged profiles derived in Section \ref{peaks} assumed purely accreting haloes, they hold for all haloes regardless of their assembly history \citep{SM19}. In fact, this remarkable result holds for all halo properties except for those related with stripped subhaloes (see Paper II). In particular, it holds for the properties of substructure regarding {\it accreted} subhaloes. Therefore, to derive the radial distribution and MF of accreted subhaloes we can assume, with no loss of generality, that their host haloes evolve by pure accretion. For simplicity in the notation, we drop from now on the arguments $M\h$ and $t\h$ referring to the final host halo in all subhalo properties.

Given the inside-out growth of accreting haloes, the mean cumulative abundance of subhaloes per infinitesimal mass around $\clM$ inside the radius $r$, ${\cal N}(<\! r, \clM)$, of the halo with mass $M(r)$ coincides with the mean cumulative abundance of nested peaks inside the corresponding peak (eq.~ [\ref{peaknum}] with $S'$ replaced by the inverse of the mean peak trajectory $\delta(S)$ tracing the halo growth by the time $t(r)$ when the halo had the mass $M(r)$). Thus the mean abundance of subhaloes per infinitesimal mass and radius around $\clM$ and $r$ is related to the mean conditional number of peaks with $\delta$ per infinitesimal mass around $\clM$ subject to being directly nested in non-nested peaks with $\delta$ at $S'$ (eq.~[\ref{peaknum}]) through
\beq
{\cal N}(r,\clM)=\frac{\partial \clS}{\partial \clM}\frac{\der {\cal N}[\clS,\delta|S(\delta),\delta]}{\der
  \delta}\frac{\der\delta}{\der S}\frac{\partial S}{\partial M}\frac{\der M}{\der r}
\label{densclump} 
\eeq
where $\delta(S)$, $S(M,t)$ and $M(r)$ are given by equations (\ref{dmd}), (\ref{rm}) and (\ref{vir2}), respectively.

The above mentioned separability of ${\cal N}[\clS,\delta|S(\delta),\delta]$ is preserved through the change of variable from $\clS$ to $\clM$ and differentiation with respect to $\delta$, so ${\cal N}(r,\clM)$ factorises (to less than 4\% error) into a function of $r$ proportional to $4\pi r^2 \rho(r)$ times a function of $\clM$, which must be equal to the subhalo MF of the halo with $M(r)$ at $t(r)$, ${\cal N}_{[M(r),t(r)]}(\clM)$. Thus the abundance of subhaloes with $\clM$ at $r$ of the halo with $M\h$ at $t\h$ is {\it very approximately} given by
\beq
{\cal N}(r,\clM)= 4 \pi\,r^2 \frac{\rho(r)}{M\h}\,{\cal N}(\clM).
\label{calN}
\eeq
More concretely, for any given $\clM$ equation (\ref{calN}) holds at $r$ such that $\clM<M(r)/3$. Otherwise, ${\cal N}(r,\clM)$ vanishes (see below).

On the other hand, as the density of the halo at $r$ is the sum of the densities of dDM and subhaloes of all masses, the dDM mass fraction at $r$ is given by
\beq
f\dc(r)=1-\frac{1}{4\pi r^2\rho(r)}\int_{M\res}^{M\h} \der \clM\,\clM\,{\cal N}(r,\clM).
\label{fdc}
\eeq

\begin{figure}
\centerline{\includegraphics[scale=.45,bb=18 50 560 558]{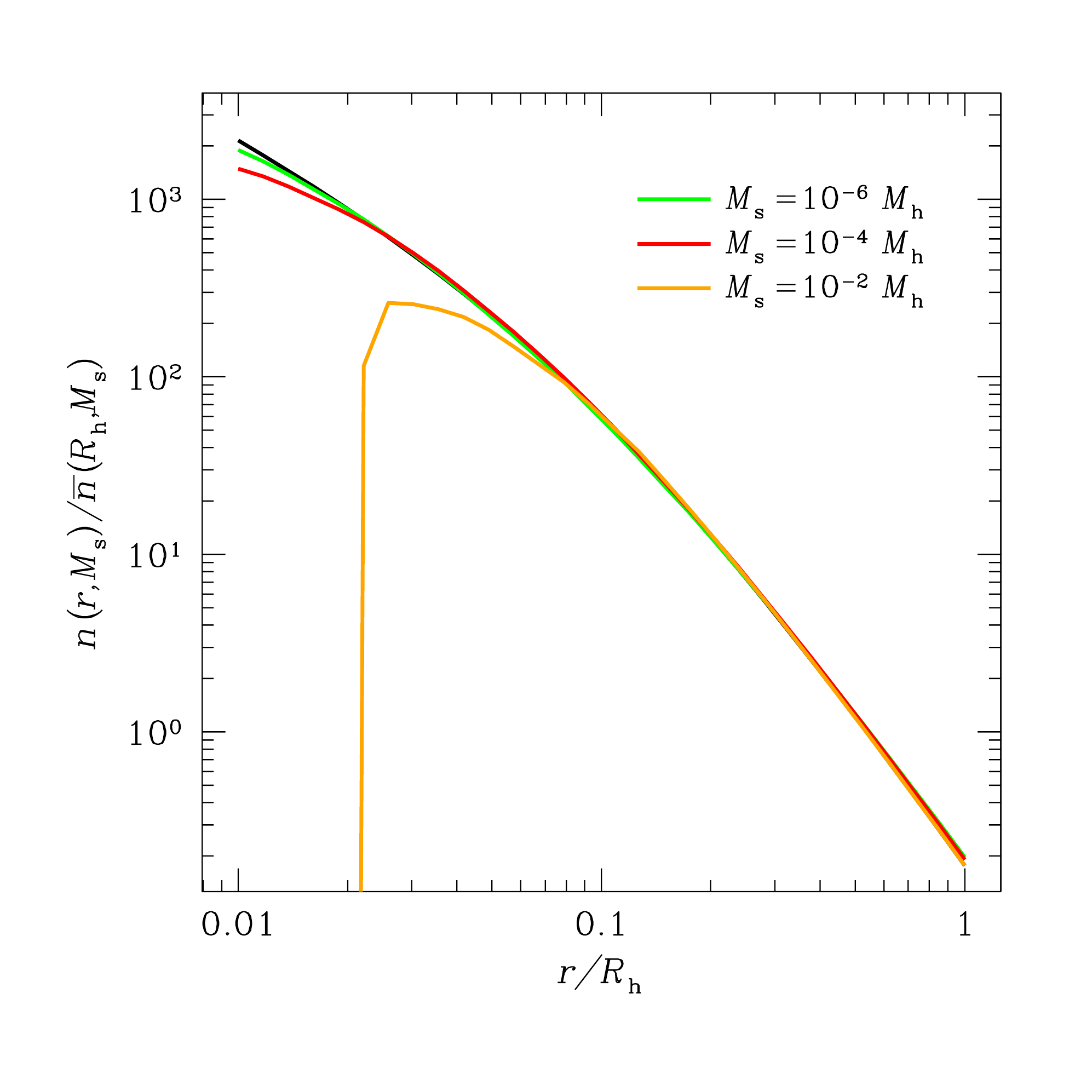}}
\caption{Scaled spherically averaged number density profiles of accreted subhaloes with several masses (coloured lines). For comparison we plot the scaled spherically averaged density profile of the host halo of the NFW form \citep{NFW97} (black line).}
(A colour version of this Figure is available in the online journal.)
\label{f3}
\end{figure}

The form of ${\cal N}(r,\clM)$ given by equation (\ref{calN}) implies that the mean spherically averaged number density profile of accreted subhaloes per infinitesimal mass around $\clM>M\res$, $n(r,\clM)={\cal N}(r,\clM)/(4\pi r^2)$, scaled to the corresponding total mean number density in the halo, $\bar n(R\h,\clM)={\cal N}(\clM)/(4\pi R\h^3/3)$, is {\it very approximately} equal to
\beq
\frac{n(r,\clM)}{\bar n(R\h,\clM)}= \frac{\rho(r)}{\bar\rho(R\h)}.
\label{new}
\eeq
The scaled number density profiles obtained {\it directly from equation} (\ref{densclump}) for several subhalo masses are plotted in Figure \ref{f3}. From now on, a bar on a function of $r$ denotes its mean value within that radius. The result is interesting in the two following respects.

First, the scaled subhalo number densities for different $\clM$ overlap as stated in HCFJ condition 1, but for a cutoff at the radius $r$ where each mass $\clM$ equals $M(r)/3$. This cutoff is not unsurprising; it reflects the fact that subhaloes of any given mass $\clM$ cannot be accreted by the host halo when its mass was $M(r)$. Indeed, that accretion would become a major merger and both the subhalo and the host halo would be destroyed (see more details below). Of course, those cutoffs will only be observed in the radial distribution of {\it accreted} subhaloes, not {\it unevolved} ones who can migrate towards smaller radii due to the effects of dynamical friction.

Second, the scaled subhalo number density profiles always follow the scaled density profile of the halo, $\rho(r)$, independently of the dDM content in the Universe or simulation, in agreement with HCFJ condition 1. Note, however, that since the fraction of putative subhaloes with masses below $M\res$ increases with decreasing radius the dDM mass fraction also does until becoming unity at some small radius. Of course, dynamical friction will also alter this property of accreted subhaloes. Indeed, as shown by HCFJ, the scaled number density profiles of unevolved subhaloes are actually not proportional to $\rho(r)$, but cuspier. The more massive subhaloes, the cuspier their number density profile. HCFJ interpreted that trend as due to the effects of dynamical friction, which causes massive subhaloes to migrate inwards. Indeed, the number density profiles of subhaloes less affected by dynamical friction were found to be roughly proportional to the halo density profile, suggesting that, if dynamical friction could be corrected, condition 1 would be satisfied. But the possibility remained that, after correcting for dynamical friction, the number density profile of unevolved subhaloes turned out to be even shallower than $\rho(r)$. In fact, as can be seen in Figures 4 and 5 of HCFJ, the number density profiles of `resolved' subhaloes (i.e. those less stripped off because having been accreted more recently) seemed to be slightly shallower than $\rho(r)$. The results of CUSP allow us to discard that possibility so that HCFJ's condition 1 is confirmed (see also \citealt{Hea18}).

Integrating over $r$ the abundance ${\cal N}(r,\clM)$ per infinitesimal mass and radius, we are led to the differential subhalo MF, ${\cal N}(\clM)$. In principle, this predicted MF is not directly comparable to the universal MF of unevolved subhaloes derived by \citet{Hea18} for simulated haloes of different masses. First, these authors dealt with subhaloes seen at the infall time (i.e. the time of first halo crossing) where their mass is maximum, while our MF refers to subhaloes at their accretion time (i.e. the time of orbit stabilisation) after having been somewhat stripped during virialisation (see Paper II). Second and more importantly, the universal MF inferred by \citet{Hea18} refers to subhaloes of all levels, while our predictions refer to first-level subhaloes only. Both differences explain that, contrarily to what happens with our MF, the integral from zero to unity of the universal MF of unevolved subhaloes provided by \citet{Hea18}, $\der {\cal N}(>m)/\der \ln (m)$, where $m$ is the subhalo mass scaled to the total mass of the halo (see their eq.~4), is not equal to 1 but to 2.15. However, stripping (during virialisation and after it) is little dependent on subhalo mass (see Paper II) and the MF of subhaloes of any higher-level should behave as the MF of first-level subhaloes in low-mass haloes in the past, i.e. they should also be essentially a power law with (essentially) the same index (see the dependence of the MF on halo mass below). Consequently, our predicted MF should still match the universal one derived by \citet{Hea18} provided it is conveniently renormalised. 

\begin{figure}
\centerline{\includegraphics[scale=.45,bb= 18 40 560 550]{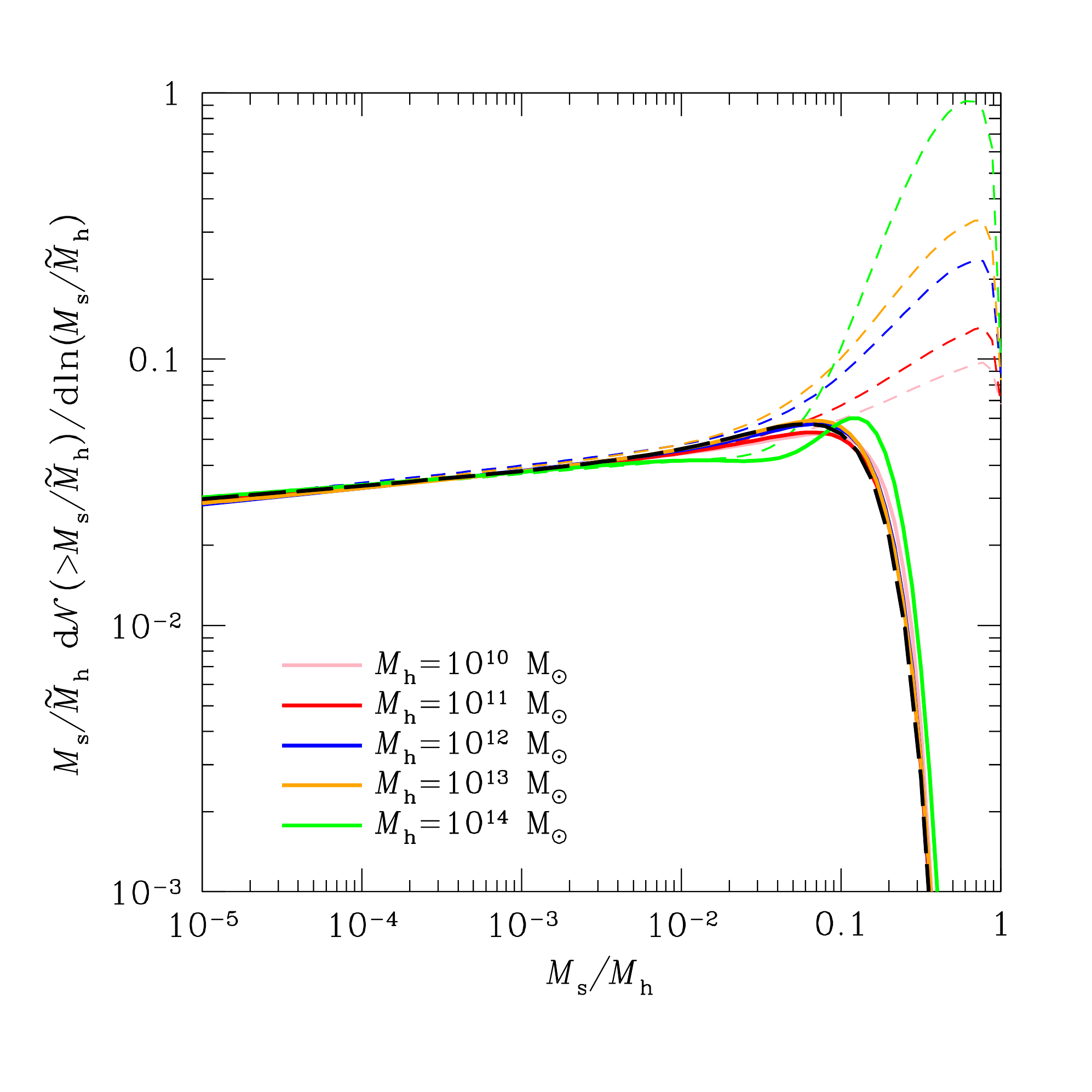}}
\caption{Differential MF of accreted subhaloes near the high-mass end predicted by CUSP for current purely accreting haloes of several masses (coloured dashed lines) and their phenomenological correction at the high mass end (coloured solid lines) accounting for the peak-peak correlation. For comparison we plot the empirical universal MF found from simulations by \citealt{Hea18} renormalised so as to correspond to real accreted subhaloes (long-dashed black line). See the text for the definition of $\tilde M\h$.}
(A colour version of this Figure is available in the online journal.)
\label{f4}
\end{figure}

As can be seen in Figure \ref{f4}, except at the very high-mass end, the predicted MFs for haloes of all masses essentially overlap with each other in agreement with the results obtained by \citet{Hea18} (see also \citealt{Gi08}) as well as with the (renormalised) universal MF derived by them. Specifically, they are close to power-laws with logarithmic slopes, $\der {\cal N}(>m)/\der \ln m$, approximately equal to $\sim -0.95$ as found by HCFJ (see also \citealt{Gi08}). To avoid any spurious effect of the predicted MFs at the high-mass end or in the extrapolation of the empirical MFs below $M\res$ the MFs plotted in Figure \ref{f4} have been scaled to $\tilde M\h$ equal to the mass of subhaloes between $10^{-5}M\h$ and $10^{-3}M\h$. Of course, if the MFs of haloes of different masses really overlap, they must do so in that particular range. This result thus shows that HCFJ condition 2 is the consequence that the function of $S$ into which the $\delta$-derivative of the mean conditional number density of nested peaks factorises is very nearly of the power-law form with the appropriate index. More exactly, the slope we find at $10^{-5}M\h$ slightly decreases from $-0.961$ to $-0.941$ for halo masses ranging from $10^{10}$ \modot to $10^{12}$ \modot and then begins to increase again, being equal to $-0.947$ and $-0.955$ for haloes with $10^{13}$ \modot and $10^{14}$ \modotc, respectively.\footnote{There is also a hint of this latter effect in the data of simulated haloes fitted by \citet{Hea18} (see the low-mass end of the MF in their Fig.~5). However, its significance cannot be assessed.} Note that as the slope is close to $-1$, even though the dependency of the asymptotic slope of the theoretical MFs on halo mass is very small, it still leads to significant differences in the mass fraction of low mass subhaloes and dDM. 

At the very high-mass end, however, the predicted MFs deviate from the empirical one. Indeed, the empirical MF shows a rapid fall off when subhalo masses approach $M\h/3$, which reflects the fact that when haloes with $M\h$ capture haloes more massive than this mass, the capture is not an accretion but a major merger causing the destruction of the subhalo (and of the capturing halo). Instead, the theoretical MFs keep on increasing beyond $M\h/3$, which is clearly meaningless. This strange behaviour of the predicted MFs is due to a deficiency in the conditional number density of nested peaks, $N\pk\nest(S,\delta|S',\delta)$, used to derive it. As explained in \citet{Mea98} that conditional number density is only approximate; it does not properly account for the peak-peak correlation at small separations, which should prevent peaks with $\delta$ at $S$ from being nested within other peaks with $\delta$ at $S'$ when $S$ approaches $S'$. Then peaks become saddle points \citep{Cea20}, which traces the major merger of the corresponding haloes. That spurious excess of subhaloes of masses tight to $M\h$ is not something to worry about: it is only transient (the excess near the halo mass at any time in the past when the halo was less massive disappears when the halo grows and a new excess appears near to the new mass of the halo) and does not affect the scaled number density derived above thanks to the separability of ${\cal N}(r,\clM)$. Nevertheless, it would be desirable to have accurate theoretical MFs up to the very high-mass end. 

Of course, the accurate derivation of the MF up to $M\h$ would require to properly account for the peak-peak correlation, which is by no means straightforward. Thus, we have opted for applying a phenomenological exponential cutoff to the predicted MF through the factor 
\beq
C(m,M\h)=A(M\h)\exp\left\{-8.9 m^{[1.9 U(M\h)]}\right\},
\label{cutoff}
\eeq
where the values of the coefficients have been taken from the empirical cutoff found by \citet{Hea18}, and the function $U(M\h)$ given by
\beq
\log U(M\h)=1+a_1 \log\left(\frac{M\h}{M_0}\right) + a_2 \left[\log\left(\frac{M\h}{M_0}\right)\right]^2,
\eeq
with $M_0=4.73\times 10^{3}$, $a_1=- 0.187$ and $a_2=0.00821$, and where $A(M\h)$ is a renormalisation constant very close to unity in all cases. After that phenomenological cutoff, all the predicted MFs essentially overlap with each other and with the empirical universal MF up the $m=1$ (see Fig.~\ref{f4}). Given the very similar behaviour of all these MFs at the low-mass end, this means that they will also coincide now for any arbitrary value of the scaling mass $\tilde M\h$ relative to $M\h$, in particular the value used in \citet{Hea18} or the whole halo mass $M\h$. Only the MF of very massive haloes ($M\h\ga 10^{14}$ M$_\odot$) somewhat deviates from the rest at the high-mass end. Whether such a deviation is due to having applied a deficient cutoff or, on the contrary, it has a real basis\footnote{The data on the most massive haloes depicted in Figure 5 of \citealt{Hea18} seem to show a hint of that deviation.} is hard to tell. Future simulations should clarify this point.

%

%
%

\begin{figure}
\centerline{\includegraphics[scale=.45,bb= 18 40 560 550]{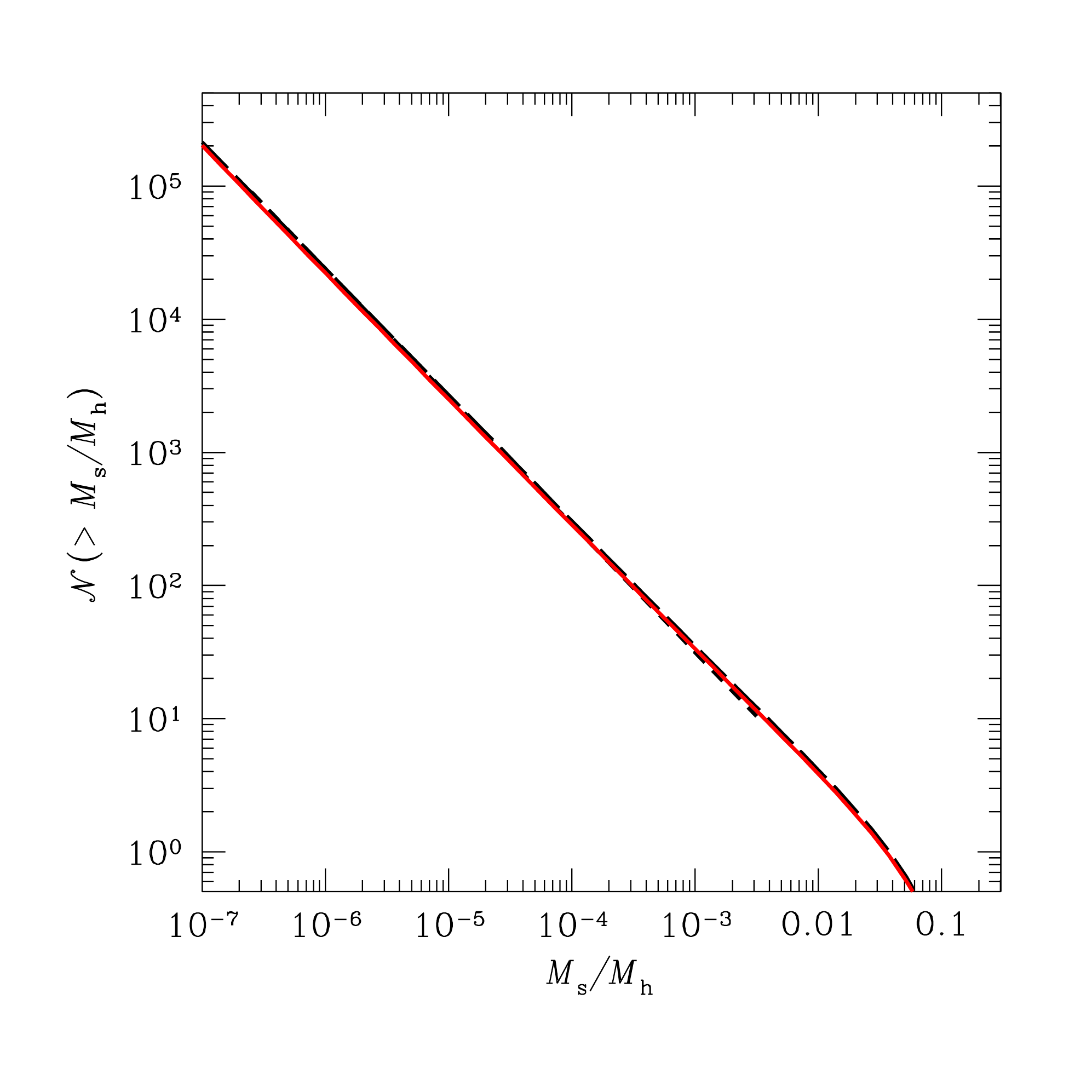}}
\caption{Cumulative MF of accreted subhaloes
  predicted by CUSP for current MW-mass haloes in Aquarius-like simulations (red solid line), compared to that found by HCFJ in the Level 1 Aquarius halo A (black short-dashed line) and by \citet{Hea18} for haloes of the same mass, in general (black long-dashed line) which is essentially hidden by the theoretical solution. Both empirical MFs have been renormalised so as to correspònd to first-level accreted subhaloes.}
  (A colour version of this Figure is available in the online journal.)
\label{f5}
\end{figure}

Lastly, integrating the (corrected) differential MF down to $\clM$, we obtain the cumulative MF ${\cal N}(> \clM)$ of accreted subhaloes in haloes with $M\h$ at $t\h$. That integral could have been directly performed from the conditional number of nested peaks (eq.~[\ref{densclump}]),
\beqa 
{\cal N}(>\clM)= \int_{\delta(t\res)}^{\delta(t\h)} \der\delta \int_{\clM}^
{M\h} \der \clM\,\frac{\der {\cal N}[\clM,\delta|S(\delta),\delta]}{\der \delta}~~~~\nonumber\\
\approx {\cal N}[>\clM,t\h|M\h,t\h],~~~~~~~~~~~~~~~~~~~~~~
\label{peaksi} 
\eeqa 
though the resulting cumulative MF should also be corrected for the spurious excess mentioned above at very large subhalo masses. To write the latter equality in equation (\ref{peaksi}) we have taken into account the separability of the integrant and that $N\{>\clM,t\res|M[r(t\res)],t\res\}$ is null because the mass of haloes inside $r(t\res)$ is smaller than $\clM$ for all relevant values of $\clM$. The cumulative MF, ${\cal N}(>\clM)$, obtained for MW-mass haloes from the previous differential one is compared in Figure \ref{f5} to that found by HCFJ in the Level 1 Aquarius halo A (after converting the mass $M_{200}$ used by these authors to $M\h$) and to the typical one for haloes of that mass found by \citet{Hea18}. As can be seen, the predicted cumulative MF is close to a power-law with index close to $-1$ (i.e. ${\cal N}(\clM)$ nearly proportional $\clM^{-2}$), although slightly larger than this according to the slopes of the differential MFs mentioned above. Note that, as the separability of the number of nested peaks is not perfect, the logarithmic slope of the predicted MF for MW-mass haloes varies from $-0.94$ at large $\clM$ to $-0.97$ at the low-mass end, to be compared with the slopes of $-0.96$ and $-0.95$ found by HCFJ and \citet{Hea18}, respectively (in the latter case at $10^{-5}M\h$). 

\section{TWO-BODY INTERACTIONS}\label{friction}

As mentioned, the small discrepancy we have found between the predictions of CUSP and the original HCFJ condition 1 seems to be due to the underestimate by those authors of the effect of dynamical friction on massive subhaloes. We cannot be more conclusive than that because we are dealing here with `accreted' subhaloes, not with `unevolved' ones, so the possible effects of two-body interactions on subhaloes orbiting within the host halo are not taken into account. But those interactions could also affect other aspects of our treatment. It is thus worthwhile discussing them.

One crucial point in CUSP is that, as justified in \citet{SM19}, we have the right to assume purely accreting haloes when deriving the properties of ordinary ones, i.e. haloes that have suffer major mergers. However, two-body relaxation causes virialised self-gravitating systems to approach energy equipartition and to suffer evaporation. Thus their inner structure must progressively adapt to these varying conditions. In the case of DM haloes, however, the timescale of two-body relaxation is much longer than the age of the Universe \citep{BT08}, so this possibility can be discarded (but see \citealt{vdB18} for the case of numerical simulations).

More importantly, two-body interactions could invalidate the conclusion reached in CUSP that accreting haloes grow inside-out. As shown in SVMS (see also \citealt{SM19}), the reason that the gentle virialisation through shell-crossing of accreting haloes leads to their inside-out growth is the particular way crossing shells exchange energy. But the gravitational energy considered in that reasoning only accounted for the {\it collective} potential well of the system (including the two crossing shells), while, when to shells cross, every object in one of the shells deflects behind itself the less massive bodies lying in the other shell due to its strong low-range gravitational pull. And the permanent overdensity of deflected small mass bodies behind the more massive deflecting one causes to it the ``extra" brake called dynamical friction. Thus dynamical friction is already active during the virialisation process itself. Nonetheless, provided deflecting objects are very numerous and, hence, uniformly distributed over the shell, the even more numerous less massive deflected bodies of the other shell will be uniformly scattered behind them and fill the whole crossed shell symmetrically, with no individual overdensity behind any deflecting object of the other shell. In other words, dynamical friction should have no noticeable effect in shell-crossing.

Only subhaloes of masses typically above a few percent of the host mass \citep{BT08} are rare enough for them not to be uniformly distributed over accreted shells and suffer significant dynamical friction. This will cause their orbits never stop contracting and the whole system to also slowly evolve. However, the mass fraction in the form of those massive objects is very small, so the dynamics of relaxed haloes is dominated by the much more numerous less massive subhaloes and DM particles, meaning that the inside-out growth of haloes during accretion should be, indeed, a good approximation. 

To sum up, the neglect of two-body interactions and dynamical friction followed here is justified because we are interested only on {\it accreted} subhaloes not on {\it unevolved} ones. Indeed, even though all subhaloes more massive than $10^{-4}M\h$ may suffer significant dynamical friction, only the most massive ones (with masses $\sim M\h/3$) may already suffer it during virialisation, i.e. during the stabilisation of their apocentres. And, as such extremely massive subhaloes are very rare, they should not affect the overall inside-out growth of accreting haloes. Of course, our predictions in Paper II drawn from the monitoring of subhalo orbits will only be valid for subhaloes with low enough masses. To overcome this shortcoming we are currently working on a new version of the present approach which includes the effects of dynamical friction. But the present version is enough for the main goal of the present series of Papers: the detailed study on the “typical” properties of low-mass subhaloes, the only ones with large enough ensembles in any individual halo \citep{Z04,Pe05,OB16}.

\section{SUMMARY}\label{dis}

We have extended the CUSP formalism, originally developed to deal with the macroscopic DM halo properties in hierarchical cosmologies, to deal with their basic components: subhaloes and dDM. To that end,  we have shown how to account in a very simple manner for the DM that cannot collapse and form haloes with masses below the free streaming mass in the case of the real Universe or the halo resolution mass in the case of a simulation in a CDM cosmology so that it stays as dDM progressively accreted onto more massive haloes. In addition, the link between haloes and non-nested peaks has been extended to subhaloes and nested peaks.  

By monitoring the accretion of subhaloes and dDM onto haloes evolving by pure accretion (monolithic collapse), we have accurately determined the abundance and radial distribution of dDM and subhaloes accreted onto haloes in two illustrative cases: a real 100 GeV WIMP universe and a typical $N$-body simulation. 

The total dDM mass fraction found in MW-mass haloes (33\% in the a 100 GeV WIMP universe and 50\% in an Aquarius-like simulation) is substantially larger than reported in previous works. This apparent discrepancy is due, however, to the fact that the only dDM counted in those works is that directly accreted onto the final halo since its last major merger, i.e. it does not count the dDM accreted onto its progenitors and transferred to the main parent halo when the progenitors merge. Of course, this amount of dDM does not include yet the extra dDM released into the intra-halo medium from stripped subhaloes (see Paper II). 

Regarding the abundance and radial distribution of accreted subhaloes (in the same wide sense above), the predictions found fully recover the results of simulations. This has allowed us to clarify the origin of HCFJ conditions 1 and 2. Specifically, we have shown that the scaled number density profiles of accreted (or unevolved) subhaloes with different masses $\clM$ overlap in one single profile, as stated in condition 1, because the derivative with respect to the density contrast $\delta$ of the number of peaks directly nested in larger scale non-nested peaks is separable in a function of scale and a function of $\delta$, which translates into a function of $r$ in the halo at any time $t$. 

In addition we have shown that the diferential MF of accreted subhaloes is very close to a power-law with logarithmic derivative $\der \ln {\cal N}(> \clM)/\der \ln \clM$ close to $-1$, as stated in HCFJ condition 2 and nearly universal, i.e. independent of halo mass, in agreement with the results of simulations \citep{Gi08,Hea18}. That behaviour is due to the particular form of the above mentioned function of peak scale into which the number of nested peaks factorises, which is nearly of the power-law form. Such a cumulative subhalo MF is independent of the amount of dDM accreted onto haloes, which only affects the very low-mass end.

All the preceding results have been obtained for purely accreting haloes, which greatly simplifies the treatment because of their inside-out growth \citep{SM19}. However, they also hold for ordinary haloes having suffered major mergers because all macroscopic properties of haloes do not depend on their assembly history (see \citealt{SM19} for a formal proof). The only exception to this general rule concerns the properties somehow related to the tidal stripping of subhaloes (see Paper II). 

The abundance and radial distribution of accreted dDM and subhaloes derived here are used, in Paper II, as initial conditions for their evolution and stripping inside the host haloes, which clarifies the origin of HCFJ condition 3. In Paper III we analyse the dependence of the resulting substructure on halo mass and halo formation time.

\par\vspace{0.75cm}\noindent
{\bf DATA AVAILABILITY}\par\vspace{7pt}\noindent 
The data underlying this article will be shared on reasonable request to the corresponding author.

\par\vspace{0.75cm}\noindent
{\bf ACKNOWLEDGEMENTS}\par\vspace{7pt}\noindent 
One of us, I.B., has benefited of a MEXT scholarship by the Japanese MECSST. Funding for this work was provided by the Spanish MINECO under projects CEX2019-000918-M of ICCUB (Unidad de Excelencia `Mar\'ia de Maeztu') and PID2019-109361GB-100 (this latter co-funded with FEDER funds) and the Catalan DEC grant 2017SGR643.

{}

\end{document}